\documentclass[12pt]{article}
\pdfoutput=1

\usepackage{amsmath,amssymb}
\usepackage{graphicx}

\makeatletter

\renewcommand\section{\@startsection{section}{1}{\z@}
                                   {-3.5ex \@plus -1ex \@minus -.2ex}
                                   {2.3ex \@plus .2ex}
                                   {\normalfont\large\bfseries}}
\renewcommand\subsection{\@startsection{subsection}{2}{\z@}
                                   {-3.25ex\@plus -1ex \@minus -.2ex}
                                   {1.5ex \@plus .2ex}
                                   {\normalfont\normalsize\bfseries}}
\renewcommand\subsubsection{\@startsection{subsubsection}{3}{\z@}
                                   {-3.25ex\@plus -1ex \@minus -.2ex}
                                   {1.5ex \@plus .2ex}
                                   {\normalfont\normalsize\bfseries}}
\renewcommand\paragraph{\@startsection{paragraph}{4}{\z@}
                                   {3.25ex \@plus1ex \@minus.2ex}
                                   {-1em}
                                   {\normalfont\normalsize\bfseries}}

\makeatother

\newcommand{\beq}{\begin{equation}}
\newcommand{\eeq}{\end{equation}}
\newcommand{\bea}{\begin{eqnarray}}
\newcommand{\eea}{\end{eqnarray}}

\newcommand{\SU}{{\rm SU}}
\newcommand{\SO}{{\rm SO}}

\newcommand{\so}{{\rm so}}

\newcommand{\C}{\mathbb C}

\newcommand{\R}{\mathbb R}

\newcommand{\id}{\hbox{1\kern-.27em l}}

\newcommand{\Tr}{{\rm Tr}}

\newcommand{\ad}{{\rm ad}}

\newcommand{\cO}{{\cal O}}

\begin{document}

\pagestyle{empty}

\begin{center}

\vspace*{30mm}
{\Large Boundary conditions for GL-twisted $N = 4$ SYM}

\vspace*{30mm}
{\large M{\aa}ns Henningson}

\vspace*{5mm}
Department of Fundamental Physics\\
Chalmers University of Technology\\
S-412 96 G\"oteborg, Sweden\\[3mm]
{\tt mans@chalmers.se}     
     
\vspace*{30mm}{\bf Abstract:} 
\end{center}
We consider topologically twisted $N = 4$ supersymmetric Yang-Mills theory on a four-manifold of the form $V = W \times\R_+$ or $V = W \times I$, where $W$ is a Riemannian three-manifold. Different kinds of boundary conditions apply at infinity or at finite distance. We verify that each of these conditions defines a `middle-dimensional' subspace of the space of all bulk solutions. Taking the two boundaries of $V$ into account should thus generically give a discrete set of solutions. We explicitly find the spherically symmetric solutions when $W = S^3$ endowed with the standard metric. For widely separated boundaries, these consist of a pair of solutions which coincide for a certain critical value of the boundary separation and disappear for even smaller separations.

\newpage \pagestyle{plain}

\section{Introduction}
Maximally supersymmetric Yang-Mills theory in four dimensions admits three inequivalent topological twistings \cite{Yamron, Vafa-Witten}. One of these, called the GL-twist in \cite{Kapustin-Witten}, plays a prominent role in applications of gauge theory to the geometric Langlands program. It leads to an elliptic set of localization equations of the form
\bea \label{first}
(F - \phi \wedge \phi + t d_A \phi)^+ & = & 0 \cr
(F - \phi \wedge \phi - t^{-1} d_A \phi)^- & = & 0 \cr
d_A (* \phi) & = & 0
\eea
for some complex parameter $t$ together with
\bea \label{second}
d_A \sigma & = & 0 \cr
[ \phi, \sigma] & = & 0 \cr
[ \sigma, \overline{\sigma} ] & = & 0 .
\eea
The notation is as follows: $d_A$ is the covariant exterior derivative associated to a connection $A$ with field strength $F = d A + A \wedge A$ on the gauge bundle $E$. (This is a principal $G$-bundle over the four-manifold $V$ on which the theory with gauge group $G$ is defined.) The other bosonic fields are a one-form $\phi$ and a complex zero-form $\sigma$ with values in the vector bundle $\ad (E)$ associated to $E$ via the adjoint representation of $G$. (The latter fields originate from the six real scalar fields of the untwisted theory.) There is a Lie product understood in the $\phi \wedge \phi$ terms. $*$ denotes the Hodge duality operator induced from the Riemannian structure on $V$, and ${}^+$ and ${}^-$ denotes the projection on self-dual and anti self-dual two-forms respectively.

As explained in \cite{Witten2011}, for applications to theories analogous to the $D3{-}D5$ system\footnote{The gauge group for the $D3{-}D5$ system was called $G^\vee$ rather than $G$ in \cite{Witten2011}.} (which by $S$-duality is related to theories analogous to the $D3{-}NS5$ system), the relevant value of the parameter $t$ is $t = +1$. (The $D3{-}\overline{D5}$ system would give $t = - 1$.) The first set of equations (\ref{first}) then simplifies to
\bea \label{d=4_equations}
F - \phi \wedge \phi + * d_A \phi & = & 0 \cr
d_A (* \phi) & = & 0 .
\eea
The second set of equations (\ref{second}) typically enforces $\sigma$ to vanish identically, and will not be considered further in this note.

The above equations can be considered on any four-manifold $V$. But following \cite{Witten2011} (see also \cite{Witten2010a, Witten2010b}), we will restrict our attention to the cases 
\beq \label{product_RW}
V = W \times \R_+ 
\eeq
and
\beq \label{product_IW}
V = W \times I  .
\eeq
Here $W$ is a (compact or non-compact) Riemannian three-manifold, the half-line $\R_+$ or the interval $I$ with linear coordinate $y$ is endowed with the standard metric $d s^2 = d y^2$, and $V$ is given the product metric. We will assume that the gauge bundle $E$, the connection $A$ and the one-form $\phi$ over $V$ are pullbacks of objects over $W$ which we, by a slight abuse of notation, denote by the same symbols. Since $\R_+$ or $I$ is contractible, this assumption about $E$ is not really a restriction, and the assumption about $A$ can be seen as a gauge condition. Finally, the assumption about $\phi$ actually follows from a vanishing theorem that can be derived from the equations (\ref{d=4_equations}) applied to $V$ of the form (\ref{product_RW}) or (\ref{product_IW}) together with some additional postulate about the boundary behaviour \cite{Kapustin-Witten, Witten2011, Witten2010b}. So henceforth $E$ is a principal $G$-bundle over $W$ with connection $A$,
\bea
F & \in & \Omega^2 (W, \ad (E)) \cr
\phi & \in & \Omega^1 (W, \ad (E)) ,
\eea
and the equations (\ref{d=4_equations}) take the form
\bea \label{d=3_equations}
\frac{\partial A}{\partial y} & = & * (d_A \phi) \cr
\frac{\partial \phi}{\partial y} & = & * (F - \phi \wedge \phi)
\eea
together with
\beq \label{moment_map}
d_A (* \phi) = 0 .
\eeq
Here $*$ is the Hodge duality operator induced from the Riemannian structure on $W$. We note that by the Bianchi identity $d_A F = 0$ and the equations (\ref{d=3_equations}), the $y$-derivative of equation (\ref{moment_map}) is identically satisfied:
\bea
\frac{\partial}{\partial y} \left( d_A (* \phi) \right) & = & d_A (* \frac{\partial \phi}{\partial y} ) + * \frac{\partial A}{\partial y} \wedge \phi - \phi \wedge * \frac{\partial A}{\partial y} \cr
& = & d_A (F - \phi \wedge \phi) + d_A \phi \wedge \phi - \phi \wedge d_A \phi \cr
& = & 0 .
\eea
So this equation need only be imposed for a single value of $y$, and will not be considered further in this note.

The space of connections $A$ is an affine space, the underlying linear space of which equals the space $\Omega^1 (W, \ad (E))$ in which $\phi$ takes its values. The solutions to the first order differential equations (\ref{d=3_equations}) can thus be parametrized by Cauchy data $(A, \phi)$ that take their values in
\beq \label{Cauchy}
\Lambda_\mathrm{Cauchy} = \Omega^1 (W, \ad (E)) \oplus  \Omega^1 (W, \ad (E))  
\eeq
on some hypersurface $y = \mathrm{const}$. The aim of this note is to further analyze two types of boundary conditions that were introduced in \cite{Witten2011}, and verify that each of them defines an essentially `middle-dimensional' subspace of the infinite dimensional space (\ref{Cauchy}). With such conditions imposed on each of the two boundary components of (\ref{product_RW}) or (\ref{product_IW}), one would thus generically expect a discrete set of solutions to (\ref{d=3_equations}).

In the next section, we consider a boundary condition that can be applied to the boundary component at infinity in (\ref{product_RW}). This states that $A + i \phi \rightarrow \rho$ as $y \rightarrow \infty$, where $\rho$ is a fixed flat connection on the complexification $E_\C$ of $E$. Up to simultaneous conjugation by an element of the complexification $G_\C$ of $G$, such a connection can be identified with a homomorphism
\beq \label{rho}
\rho \colon \pi_1 (W) \rightarrow G_\C .
\eeq

In section three, we instead consider a boundary condition that can be applied to the boundary component at finite distance in (\ref{product_RW}) or to either of the two boundary components in (\ref{product_IW}). This states that the restriction of $\ad (E)$ to a boundary component at e.g. $y = 0$ is determined by the tangent bundle $T W$ of $W$. The restriction of $E$ itself to the boundary is then determined up to to a tensor product with a homomorphism 
\beq \label{sigma}
\sigma \colon \pi_1 (W) \rightarrow C , 
\eeq
where $C$ is the center subgroup of $G$. Furthermore, $\phi = \cO (y^{-1})$ and $A = \cO (1)$ as $y \rightarrow 0$, with the residue of $\phi$ and the limiting value of $A$ determined by the Riemannian structure of $W$.

Explicit exact solutions to (\ref{d=3_equations}) with these boundary conditions can only be found for very particular choices of data (i.e. the topology and the Riemannian structure on $W$ and the homomorphisms $\rho$ and $\sigma$). The basic example when $W = \R^3$ with the standard flat metric was described in \cite{Gaiotto-Witten} following earlier work in \cite{Nahm, Diaconescu, Constable-Myers-Tafjord}. A more complicated case with a 't~Hooft operator singularity along a straight line in the boundary $W = \R^3$ was treated in \cite{Witten2011} for the case when $\mathrm{rank \;} G = 1$. In section four, we will give a (not completely) explicit description of spherically symmetric solutions (without 't~Hooft operators) for the case when $W = S^3$ with the standard round metric. For widely separated boundaries, there are two such solutions. As the boundary separation is decreased to a certain critical value, these coincide, and for even smaller separations they disappear. The two solutions that exist for a larger than critical boundary separation should therefore be connected by a five-dimensional tunneling configuration, as described in \cite{Witten2011}. This seems to imply that supersymmetry is spontaneously broken in this situation.

Some outstanding problems for which we hope that the results presented here may be useful are: To extend the analysis of \cite{Witten2011} of 't Hooft operators in the boundary to more general cases, e.g. $\mathrm{rank \;} G \geq 2$, t Hooft operators inserted along a not necessarily geodesic curve in a possibly curved $W$, or operators with non-trivial values of the monodromy parameters. One may also study tunneling between different solutions. Finally, it would be highly desirable to get a better understanding of the roles of the homomorphisms $\rho$ and $\sigma$, especially in connection with electric-magnetic $S$-duality.

\section{Boundary conditions at infinity}
As mentioned in the introduction, the boundary conditions at infinity on $V = W \times \R_+$ can be specified by a flat connection $\rho$ on the complexification $E_\C$ of $E$, which can be identified with a homomorphism (\ref{rho}). We can decompose $\rho$ into its real and imaginary parts
\beq
\rho = \rho_1 + i \rho_2 , 
\eeq
where $\rho_1$ is a connection on $E$ and $\rho_2 \in \Omega^1 (W, \ad (E))$. Flatness means that
\bea
0 & = & d \rho + \rho \wedge \rho \cr
& = & f_1 - \rho_2 \wedge \rho_2 + i d_{\rho_1} \rho_2 ,
\eea
where $f_1 = d \rho_1 + \rho_1 \wedge \rho_1 \in \Omega^2 (W, \ad (E))$ is the curvature of $\rho_1$.

If we now expand $A$ and $\phi$ around $\rho_1$ and $\rho_2$ respectively as 
\bea
A & = & \rho_1 + a \cr
\phi & = & \rho_2 + \varphi ,
\eea
the boundary conditions can be stated in terms of $a, \varphi \in \Omega^1 (W, \ad (E))$:
\beq \label{infinite_conditions}
\left( \begin{matrix} a \cr \varphi \end{matrix} \right) \rightarrow \left( \begin{matrix} 0 \cr 0 \end{matrix} \right) \;\;\;\;\; \mathrm{as} \;\;\;\;\; y \rightarrow \infty .
\eeq

To solve the equations (\ref{d=3_equations}) with these boundary conditions, we begin by writing them in the form
\beq \label{infinite_expansion}
\frac{\partial}{\partial y} \left( \begin{matrix} a \cr \varphi \end{matrix} \right) = \left( \begin{matrix} K & L \cr L & -K \end{matrix} \right) \left( \begin{matrix} a \cr \varphi \end{matrix} \right) + \left( \begin{matrix} a \wedge \varphi + \varphi \wedge a \cr a \wedge a - \varphi \wedge \varphi \end{matrix} \right) ,
\eeq
where we have introduced the linear operators
\bea
K = * ( \rho_2 \wedge . + . \wedge \rho_2)  & \colon & \Omega^1 (W, \ad (E)) \rightarrow \Omega^1 (W, \ad (E)) \cr
L = * d_{\rho_1} & \colon & \Omega^1 (W, \ad (E)) \rightarrow \Omega^1 (W, \ad (E)) .
\eea

We will need to understand the spectrum of the operator-valued $2 \times 2$-matrix 
\beq
M = \left( \begin{matrix} K & L \cr L & -K \end{matrix} \right) . 
\eeq
The operators $K$ and $L$ are symmetric with respect to the real inner product
\beq
\eta \cdot \eta^\prime = \int_W \Tr \left( * \eta \wedge \eta^\prime \right)
\eeq
on $\Omega^1 (W, \ad (E))$, so $M$ is symmetric with respect to the real inner product
\beq
\left( \begin{matrix} a \cr \varphi \end{matrix} \right) \cdot \left( \begin{matrix} a^\prime \cr \varphi^\prime \end{matrix} \right) = a \cdot a^\prime + \varphi \cdot \varphi^\prime
\eeq
and thus has a real spectrum. Furthermore, the similarity transformation
\beq
S M S^t = - M
\eeq
with 
\beq
S = \left( \begin{matrix} 0 & \id \cr - \id & 0 \end{matrix} \right)
\eeq
shows that this spectrum is symmetric around zero.
For simplicity we assume the spectrum to be purely discrete with eigenvalues $-\lambda_i$ and $+\lambda_i$, where $i$ takes its values in some (infinite) discrete set and $\lambda_i > 0$. (Zero eigenvalues correspond to first order deformations of the flat connection $\rho$, and thus change the boundary condition.) We denote the corresponding projections of an arbitrary vector $\left( \begin{matrix} a \cr \varphi \end{matrix} \right)$ as $\left( \begin{matrix} a \cr \varphi \end{matrix} \right)_i^-$ and $\left( \begin{matrix} a \cr \varphi \end{matrix} \right)_i^+$  respectively. 

The equation (\ref{infinite_expansion}) can now be written as a set of integral equations:
 \bea
 \left( \begin{matrix} a \cr \varphi \end{matrix} \right)^+_i & = & - e^{\lambda_i y} \int_y^\infty dy e^{- \lambda_i y} \left( \begin{matrix} a \wedge \varphi + \varphi \wedge a \cr a \wedge a - \varphi \wedge \varphi \end{matrix} \right)^+_i \cr
 \left( \begin{matrix} a \cr \varphi \end{matrix} \right)^-_i & = & e^{- \lambda_i y} \int dy e^{\lambda_i y} \left( \begin{matrix} a \wedge \varphi + \varphi \wedge a \cr a \wedge a - \varphi \wedge \varphi \end{matrix} \right)^-_i .
\eea
In the first equation, where the integral is multiplied by a prefactor that grows exponentially as $y \rightarrow \infty$, the integral must be taken over the indicated domain to comply with the boundary conditions (\ref{infinite_conditions}).  But in the second equation, where the prefactor is exponentially decaying, the indefinite integral is defined only up to an additive constant. The equations can now be solved recursively, starting with the conditions
\bea
\left( \begin{matrix} a \cr \varphi \end{matrix} \right)^+_i & = & \cO (e^{- 2 \lambda_\mathrm{min} y }) \cr
\left( \begin{matrix} a \cr \varphi \end{matrix} \right)^-_i & = & e^{- \lambda_i y} \left( \begin{matrix} \alpha \cr \beta \end{matrix} \right)_i^- + \cO (e^{- 2 \lambda_\mathrm{min} y })
\eea
Here $\lambda_\mathrm{min}$ is the smallest of the $\lambda_i$, and $\left( \begin{matrix} \alpha \cr \beta \end{matrix} \right)$ are arbitrary integration constants that we can require to fulfill the projection conditions
\beq
\left( \begin{matrix} \alpha \cr \beta \end{matrix} \right)_i^+ = 0 .
\eeq
So this type of boundary condition indeed defines a `middle-dimensional' subspace of the space of all bulk solutions to (\ref{d=3_equations}) parametrized by $\Lambda_\mathrm{Cauchy}$ in (\ref{Cauchy}). The general solution is given by polynomials in $y$ multiplied by exponential suppression factors. 

We remark that the result that the boundary condition defines a `middle-dimensional' subspace is derived in a simpler way in \cite{Witten2010a, Witten2010b} by considering the real part of a complex multiple of the Chern-Simons functional for the $E_\C$ connection $A + i \phi$ as a Morse function\footnote{I thank E.~Witten for a clarifying comment on this point.}. The present analysis gives a somewhat more detailed picture of the solutions, though.

 \section{Boundary conditions at finite distance}
 Before we can describe the boundary conditions at a finite value of $y$, which we take to be $y = 0$, we review some Lie algebra theory following \cite{Witten2011}: Let
\beq
\xi \colon \so (3) \hookrightarrow g ,
\eeq
where $g$ is the Lie algebra of $G$, be a principal embedding \cite{Kostant}. Under the adjoint action of this copy of $A_1 = \so (3)$ (i.e. the image of $\xi$), $g$ decomposes as
\beq \label{g-decomposition}
g = V_{j_1} \oplus \ldots \oplus V_{j_r} .
\eeq
Here $r$ is the rank of $g$, the integers (known as the `exponents' of the Lie algebra)
\beq
1 = j_1 \leq \ldots \leq j_r
\eeq
 are given by the orders of the independent Casimir operators on $g$ minus $1$, and $V_j$ denotes the $(2 j + 1)$-dimensional `spin $j$' representation of $\so (3)$ (i.e. the traceless symmetric rank $j$ tensors in three dimensions). See table 1 for the dimensions and exponents of the simple Lie algebras. 
\begin{table}
$$
\begin{array}{lll}
g & \dim g & j_1, \ldots, j_r \cr
\hline
A_r & r^2 + 2 r & 1, \ldots, r \cr
B_r & 2 r^2 + r & 1, 3, \ldots, 2 r - 1 \cr 
C_r & 2 r^2 + r & 1, 3, \ldots, 2 r - 1 \cr
D_r & 2 r^2 - r & 1, 3, \ldots, 2 r - 3, r -1 \cr
E_6 & 78 & 1, 4, 5, 7, 8, 11 \cr
E_7 & 133 & 1, 5, 7, 9, 11, 13, 17 \cr
E_8 & 248 & 1, 7, 11, 13, 17, 19, 23, 29 \cr
F_4 & 52 & 1, 5, 7, 11 \cr
G_2 & 14 & 1, 5 .
\end{array}
$$
\caption{Dimensions and exponents of simple Lie algebras}
\end{table}

The decomposition (\ref{g-decomposition}) of the Lie algebra $g$ gives rise to an analogous decomposition of $\ad (E)$ as a direct sum of vector bundles $E_{j_1}, \ldots, E_{j_r}$ associated to the frame bundle of $W$ via the $\SO (3)$ representations $V_{j_1}, \ldots, V_{j_r}$. It follows that
\beq \label{ad-decomposition}
\Omega^1 (W, \ad (E)) = \Omega^1 (W, E_{j_1}) \oplus \ldots \oplus \Omega^1 (W, E_{j_r}) .
\eeq
We denote the corresponding projections of an arbitrary section $\eta \in \Omega^1 (W, \ad (E))$ as $\eta_{j_1}, \ldots, \eta_{j_r}$.
As remarked in the introduction, this description of $\ad (E)$ does not quite determine the principal bundle $E$ itself; different choices are related by tensoring with a flat bundle that can be identified with a homomorphism as in (\ref{sigma}). 

Since $j_1 = 1$, the first term $E_{j_1} = E_1$ in the decomposition of $\ad (E)$ is isomorphic to the tangent bundle $T W$ of $W$, i.e. to the vector bundle associated to the tangent frame bundle of $W$ via the adjoint representation of $\SO (3)$. The isomorphism
\beq
e \colon T W \rightarrow E_1 
\eeq
can be regarded as a choice of vielbein on $W$:
\beq
e \in \Omega^1 (W, E_1) .
\eeq
It obeys
\beq
* (e \wedge e) = e ,
\eeq
where an $\so (3)$ Lie algebra product is understood on the left hand side. It also obeys
\beq
d_\omega e = 0 ,
\eeq
where $d_\omega$ is the covariant exterior derivative associated to the Riemannian (spin) connection $\omega$ on $W$.

If we now expand $A$ and $\phi$ around $\omega$ and $y^{-1} e$ as
\bea
A & = & \omega + a \cr
\phi & = & y^{-1} e + \varphi 
\eea
with 
\beq
a, \varphi \in \Omega^1 (W, \ad (E)) ,
\eeq
the boundary conditions at $y = 0$ are that $a = \cO (y^\epsilon)$ and $\varphi = \cO (y^{-1 + \epsilon})$ for some $\epsilon > 0$. As we will see below, these conditions in fact imply that
\bea \label{y=0_conditions}
a & = & \cO (y) \cr
\varphi & = & \cO (y \log y) .
\eea

We have
\bea
d_A \phi & = & y^{-1} (e \wedge a + a \wedge e) + d_\omega \varphi + (a \wedge \varphi + \varphi \wedge a) \cr
\phi \wedge \phi & = & y^{-2} * e + y^{-1} (e \wedge \varphi + \varphi \wedge e) + \varphi \wedge \varphi \cr
F & = & R + d_\omega a + a \wedge a ,
\eea
where in the last equation
\beq
R = d \omega + \omega \wedge \omega \in \Omega^2 (W, E_1) \subset \Omega^2 (W, \ad (E))
\eeq
is the Riemannian curvature of $\omega$. The equations (\ref{d=3_equations}) can now be written as
\bea \label{y=0_equations}
\frac{\partial a}{\partial y} & = & * \left( y^{-1} (e \wedge a + a \wedge e) + d_\omega \varphi + a \wedge \varphi + \varphi \wedge a \right) \cr
\frac{\partial \varphi}{\partial y} & = & * \left( - y^{-1} (e \wedge \varphi + \varphi \wedge e) + R + d_\omega a + a \wedge a - \varphi \wedge \varphi \right) .
\eea

To solve these equations, we begin by discussing a further decomposition of the terms in (\ref{ad-decomposition}). By (the dual of) the isomorphism $e$, we have
\bea \label{Clebsch-Gordan}
\Omega^1 (W, E_j) & \simeq & \Omega^0 (W, E_j \otimes T^* W) \cr
& \simeq & \Omega^0 (W, E_j \otimes E_1) \cr
& \simeq & \Omega^0 (W, E_j^{-} \oplus E_j^0 \oplus E_j^{+}) \cr
& \simeq & \Omega^0 (W, E_j^{-}) \oplus \Omega^0 (W, E_j^{0}) \oplus \Omega^0 (W, E_j^{+}) ,
\eea
where 
\bea
E_j^{-} & \simeq & E_{j - 1} \cr
E_j^{0} & \simeq & E_j \cr
E_j^{+} & \simeq & E_{j + 1}
\eea
by the $\so (3)$ Clebsch-Gordan series. We denote the projections of an arbitrary section $\eta_j \in \Omega^1 (W, E_j)$ on the subspaces on the right hand side as $\eta_j^{-}$, $\eta_j^{0}$, and $\eta_j^{+}$ respectively. In particular, the decomposition of the vielbein $e \in \Omega^1 (W, E_1)$ is simply
\beq
e = e^-_1 .
\eeq
We can also decompose the dual $* R \in \Omega^1 (W, E_1)$ of the Riemannian curvature:
\beq \label{Riemann-decomposition}
*R = (*R)^-_1 + (*R)^+_1 ,
\eeq
where the first term is the curvature scalar and the second term is the traceless part of the Ricci tensor. (Recall that these determine the Riemann tensor completely in three dimensions). The linear differential map
\beq
* d_\omega \colon \Omega^1 (W, \ad (E)) \rightarrow \Omega^1 (W, \ad (E)) 
\eeq
maps each subspace in the decomposition (\ref{ad-decomposition}) into itself. Under the more refined decomposition (\ref{Clebsch-Gordan}) it acts as
\bea
* d_\omega \colon \Omega^0 (W, E_j^{-}) & \rightarrow & \Omega^0 (W, E_j^{0}) \cr
* d_\omega \colon \Omega^0 (W, E_j^{0}) & \rightarrow & \Omega^0 (W, E_j^{-}) \oplus \Omega^0 (W, E_j^{0}) \oplus \Omega^0 (W, E_j^{+}) \cr
* d_\omega \colon \Omega^0 (W, E_j^{+}) & \rightarrow & \Omega^0 (W, E_j^{0}) \oplus \Omega^0 (W, E_j^{+}) .
\eea
The linear algebraic map 
\beq
* (e \wedge . + . \wedge e) \colon \Omega^1 (W, \ad (E)) \rightarrow \Omega^1 (W, \ad (E)) 
\eeq
respects not only (\ref{ad-decomposition}) but also (\ref{Clebsch-Gordan}). In fact, the subspaces appearing in the latter decomposition are eigenspaces to this map\footnote{
This corresponds to the map $- \sum_{a = 1}^3 T^a_j \otimes T^a_1 = \frac{1}{2} \sum_{a = 1}^3 \left(T^a_1 T^a_1 + T^a_j T^a_j - (T_j + T_1)^a (T_j + T_1)^a \right)$ acting on the tensor product representation $V_j \otimes V_1$ of $\so (3)$. Here $T^a_1$ and $T^a_j$ are the generators of the spin $1$ and spin $j$ representations respectively. The eigenvalue is thus $\frac{1}{2}(2 + j (j + 1) - J (J + 1))$ when acting on states with total spin $J =j- 1, j,j + 1$ corresponding to $E_j^-, E_j^0, E_j^+$. 
} with eigenvalues $j + 1$, $1$, and $- j$ respectively:
\bea
* (e \wedge \eta_j^{-} +  \eta_j^{-} \wedge e) & = & (j + 1) \eta_j^{-} \cr
* (e \wedge \eta_j^{0} +  \eta_j^{0} \wedge e) & = & \eta_j^{0} \cr
* (e \wedge \eta_j^{+} +  \eta_j^{+} \wedge e) & = & - j \eta_j^{+} .
\eea

Inserting these results into (\ref{y=0_equations}) gives
\bea
\frac{\partial a_j^-}{\partial y} & = & \frac{j + 1}{y} a_j^- +  \left( * d_\omega \varphi + * (a \wedge \varphi + \varphi \wedge a) \right)_j^-   \cr
\frac{\partial a_j^0}{\partial y} & = & \frac{1}{y} a_j^0 +  \left( * d_\omega \varphi + * (a \wedge \varphi + \varphi \wedge a) \right)_j^0   \cr
\frac{\partial a_j^+}{\partial y} & = & - \frac{j}{y} a_j^+ +  \left( * d_\omega \varphi + * (a \wedge \varphi + \varphi \wedge a) \right)_j^+   \cr
\frac{\partial \varphi_j^-}{\partial y} & = & - \frac{j + 1}{y} \varphi_j^- +  \left( * R + * d_\omega a + * (a \wedge a - \varphi \wedge \varphi) \right)_j^-   \cr
\frac{\partial \varphi_j^0}{\partial y} & = & - \frac{1}{y} \varphi_j^0 +  \left( * d_\omega a + * (a \wedge a - \varphi \wedge \varphi) \right)_j^0   \cr
\frac{\partial \varphi_j^+}{\partial y} & = & \frac{j}{y} \varphi_j^+ +  \left( * R + * d_\omega a + * (a \wedge a - \varphi \wedge \varphi) \right)_j^+ , 
\eea
where $(* R)_j^- = (* R)_j^+ = 0$ for $j > 1$ according to (\ref{Riemann-decomposition}). The general solution is
\bea
\varphi_j^0 & = & y^{-1} \int_0^y d y^\prime  y^{\prime} \left( * d_\omega a + * (a \wedge a - \varphi \wedge \varphi) \right)_j^0  \cr
a_j^+ & = & y^{- j} \int_0^y d y^\prime  (y^{\prime})^{j} \left( * d_\omega \varphi + * (a \wedge \varphi + \varphi \wedge a) \right)_j^+ \cr
\varphi_j^- & = & y^{- j - 1} \int_0^y d y^\prime  (y^{\prime})^{j + 1} \left( * R + * d_\omega a + * (a \wedge a - \varphi \wedge \varphi) \right)_j^- \cr
a_j^- & = & y^{j + 1} \int d y  y^{- j - 1} \left( * d_\omega \varphi + * (a \wedge \varphi + \varphi \wedge a) \right)_j^- \cr
\varphi_j^+ & = & y^{j} \int d y  y^{- j} \left( * R + * d_\omega a + * (a \wedge a - \varphi \wedge \varphi) \right)_j^+ \cr
a_j^0 & = & y \int d y  y^{- 1} \left( * d_\omega \varphi + * (a \wedge \varphi + \varphi \wedge a) \right)_j^0 .
\eea
In the first three equations, where the prefactors are given by negative powers of $y$, the integrals must be taken over the indicated domains to comply with the boundary conditions (\ref{y=0_conditions}). But in the last three equations, where the prefactors are given by positive powers of $y$, the indefinite integrals are defined only up to additive constants. 

With $a = \cO (y^\epsilon)$ and $\varphi = \cO (y^{-1 + \epsilon})$,  the right hand sides of the equations for $\varphi$ are dominated by the $\varphi \wedge \varphi$ terms. Performing the integrations gives $\varphi = \cO (-1 + 2 \epsilon)$. Repeating this procedure eventually gives the conditions (\ref{y=0_conditions}). The expansions of $a$ and $\varphi$ as power series in $y$ (generically including logarithmic terms) can now be determined recursively. Indeed, if the expressions are known up to terms of order $y^{k + \epsilon}$ for some $k \geq 0$, inserting them in the right hand side of the equations will determine them to order $y^{k + 1 + \epsilon}$. In this process, arbitrary integration constants $c_j^-$, $c_j^+$ and $c_j^0$, that can be seen as the components of an arbitrary element 
\beq
c = \sum_j c_j^- + c_j^+ + c_j^0 \in \Omega^1 (W, \ad (E)) ,
\eeq
will appear for the first time as the coefficients of $y^{j + 1}$, $y^j$ and $y$ in the expressions for $a_j^-$, $\varphi_j^+$, and $a_j^0$ respectively. So also this type of boundary condition defines a `middle dimensional' subspace of the space of all bulk solutions to (\ref{d=3_equations}) parametrized by $\Lambda_\mathrm{Cauchy}$ in (\ref{Cauchy}). 

We exemplify this procedure by giving the $j = 1$ components of $a$ and $\varphi$ including terms up to second order in $y$: 
\bea
a_1^+ & = & \left( \frac{1}{3} y^2 \log y - \frac{1}{9} y^2 \right) \left( * d_\omega (* R)^+_1 \right)^+_1 + \frac{1}{3} y^2 \left( * d_\omega c_1^+ \right)_1^+ \cr
& & + \cO (y^{3 - \epsilon}) \cr
a_1^0 & = & y c_1^0 \cr
& & + \left( y^2 \log y - y^2 \right) \left( * d_\omega (* R)_1^+ \right)_1^0 + y^2 \left( * d_\omega c_1^+ \right)_1^0 + \frac{1}{3} y^2 \left( * d_\omega (* R)^-_1 \right)_1^0 \cr
& & + \cO (y^{3 - \epsilon}) \cr
a_1^- & = & y^2 c_1^- \cr
& & + \cO (y^{3 - \epsilon}) \cr
\varphi_1^+ & = & y \log y (* R)^+_1 + y c_1^+ \cr
& & + y^2 \left(* d_\omega c_1^0 \right)_1^+ \cr
& & + \cO (y^{3 - \epsilon}) \cr
\varphi_1^0 & = & \frac{1}{3} y^2 \left(* d_\omega c_1^0 \right)_1^0 \cr
& & + \cO (y^{3 - \epsilon}) \cr
\varphi_1^- & = & \frac{1}{3} y (* R)^-_1 \cr
& & + \frac{1}{4} y^2 \left(* d_\omega c_1^0 \right)_1^- \cr
& & + \cO (y^{3 - \epsilon})
\eea 
Sofar, the results are universal in the sense that they are independent of $G$. Beyond second order, the non-linear terms in the equations, which do depend on $G$, start to become relevant. The possible values of $j > 1$ depend on $G$ as described above, but apart from that, the results up to second order in $y$ are universal also for $j > 1$ and can be readily determined. (A simplifying feature for $j > 1$ is that the Riemannian curvature terms only enter indirectly via the non-linear couplings to the $j = 1$ terms, and thus do not influence the results to first and second order in $y$.) 

 \section{Spherically symmetric solutions}
 The equations (\ref{d=3_equations}) simplify considerably in the case when $W$ is endowed with a maximally symmetric metric (which in three dimensions is equivalent to an Einstein metric). The traceless part $(* R)^+_1$ of the Ricci tensor then vanishes, and the curvature scalar $(* R)^-_1$ is a constant multiple of the vielbein $e = e^-_1$. A positive, negative or zero value of this constant corresponds to (a discrete quotient of) a sphere, a hyperbolic space, or a flat space respectively. 
 
We will consider the case when 
\beq
W = S^3 = \SU (2)
\eeq
with the standard round metric. Choosing the vielbein as the Maurer-Cartan orthonormal frame of left-invariant vector fields $e = g^{-1} d g$, we have 
\bea
d e & = & - e \wedge e \cr
\omega & = & \frac{1}{2} e 
\eea
and as always $* (e \wedge e) = e$. For any gauge group $G$, we can make the maximally symmetric Ansatz
\bea \label{Ansatz}
A & = & (1 + u) \omega \cr
\phi & = & s e ,
\eea
where $u$ and $s$ are functions of $y$ only. The equations (\ref{d=3_equations}) then read
\bea \label{ODE}
\frac{d u}{d y} & = & 2 s u \cr
\frac{d s}{d y} & = & \frac{1}{4} u^2  - s^2 - \frac{1}{4} .
\eea
The flow in the $s u$-plane is sketched in figure 1. 
\begin{figure}[h]
\centering
\resizebox{100mm}{!}{\includegraphics*[viewport = 300 300 600 500]{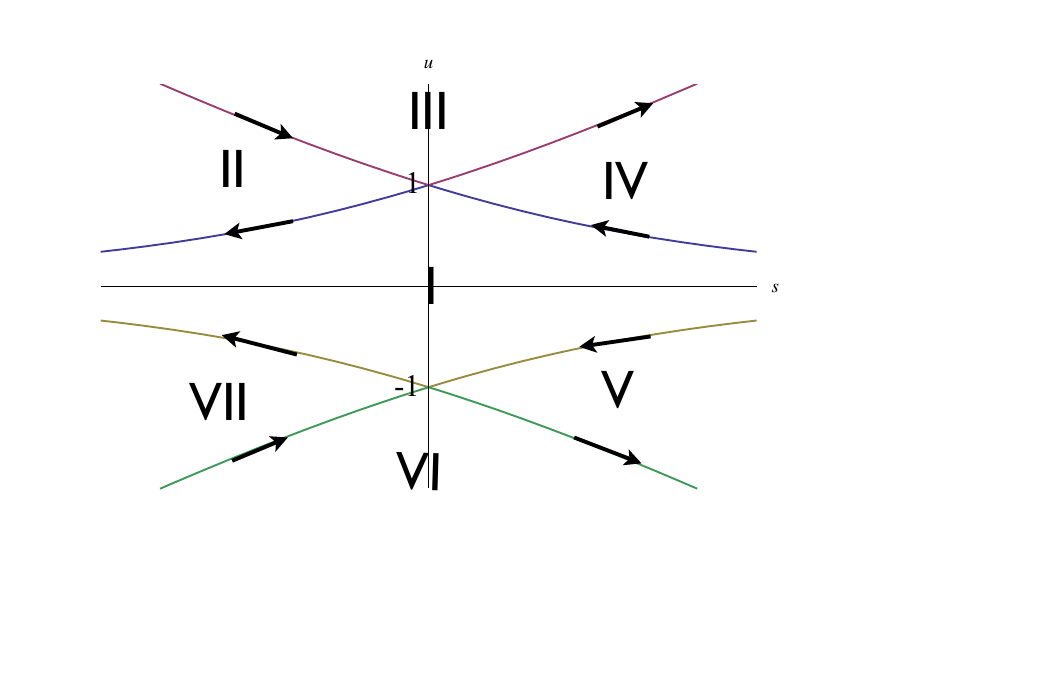}}
\caption{The flow in the $s u$-plane with arrows in the direction of increasing $y$. For clarity, only solutions flowing to or from the critical points at $s = 0$, $u = \pm 1$ have been indicated. These divide the the $s u$-plane into regions I-VII.}
\end{figure}

The two constant solutions at the critical points $s = 0, u = \pm 1$ correspond to the trivial configurations $\phi = 0$, $A = 0$ and $\phi = 0$, $A = g^{-1} d g$ respectively. These are in fact related to each other by  a `large' gauge transformation with parameter given by the identity map $g \colon W \rightarrow \SU (2)$ of unit winding number. This is the unique solution which fulfills the boundary conditions discussed in section two both for $y \rightarrow \infty$ and (with obvious modifications) $y \rightarrow - \infty$.

Translations in $y$ act freely on all other solutions to (\ref{ODE}). Solutions in (the interior of the union of) regions II-VII behave as $s, u \rightarrow \pm \infty$ with $u /s \rightarrow \pm \sqrt{12}$ as $y \rightarrow \infty$ and/or $y \rightarrow - \infty$, so they do not comply with the boundary conditions we have been discussing. Solutions in the interior of region I behave as $s \rightarrow \infty$, $u \rightarrow 0$ as $y \rightarrow y_0$ and $s \rightarrow -\infty$, $u \rightarrow 0$ as $y \rightarrow y_1$, corresponding to the boundary conditions discussed in section three both at $y = y_0$ and (with obvious modification) $y = y_1$. The distance $\Delta y = y_1 - y_0 > 0$ between the boundaries depends on the solution in question. The smallest distance is obtained for (translates of) the solution 
\bea
u & = & 0 \cr
s & = & - \frac{1}{2} \tan \frac{y}{2} ,
\eea
for which $\Delta y = \pi - (-\pi) = 2 \pi$. Solutions with arbitrary higher values of $\Delta y > 2 \pi$ appear pairwise. They are related by reflection in the $u$-axis, but are not gauge equivalent. Since these solutions disappear at $\Delta y = 2 \pi$, we expected them to be connected by a tunneling instanton solution of a set of five-dimensional equations described in \cite{Witten2011}. It would be interesting to construct this more explicitly. This result implies that supersymmetry is spontaneously broken by this configuration\footnote{unless there are further solutions which are not covered by the spherically symmetric Ansatz (\ref{Ansatz}). This seems unlikely, though.}. For the limiting case of solutions along (parts of) the boundary of region I, we have $\Delta y \rightarrow \infty$. There is then one boundary at finite $y$ (at which $s \rightarrow \pm \infty$ and $u \rightarrow 0$) and one boundary at infinity (at which $s \rightarrow 0$ and $u \rightarrow \pm 1$). 

\vspace*{5mm}
This research was supported by grants from the G\"oran Gustafsson foundation and the Swedish Research Council.

\end{document}